\documentclass[aps, pre,twocolumn,superscriptaddress,showpacs,amsmath,amssymb]{revtex4-2} 
\usepackage{amsmath}
\usepackage{amssymb}
\usepackage{dcolumn}
\usepackage{graphicx}
\usepackage{bm}
\usepackage{epstopdf}
\usepackage{threeparttable}
\usepackage{float}
\usepackage[export]{adjustbox}
\usepackage{bbm}
\usepackage{xcolor}

\begin{document}

\title{Thermodynamics of the Ising model encoded in restricted Boltzmann machines } 
 \author{Jing Gu} 
   \affiliation{Division of Natural and Applied Sciences, Duke Kunshan University, Kunshan, Jiangsu, 215300, China}
   
 \author{Kai Zhang} 
 \email{kai.zhang@dukekunshan.edu.cn}
  \affiliation{Division of Natural and Applied Sciences, Duke Kunshan University, Kunshan, Jiangsu, 215300, China}
  \affiliation{Data Science Research Center (DSRC), Duke Kunshan University, Kunshan, Jiangsu, 215300, China}
 

\begin{abstract}
The restricted Boltzmann machine (RBM) is a two-layer energy-based model that uses its hidden-visible connections to learn the underlying distribution of visible units, whose interactions are often complicated by high-order correlations. Previous studies on the Ising model of small system sizes have shown that RBMs are able to accurately learn the Boltzmann distribution and reconstruct thermal quantities at temperatures away from the critical point $T_c$.  How the RBM encodes the Boltzmann distribution and captures the phase transition are, however, not well explained. In this work, we perform  RBM learning  of the $2d$ and $3d$ Ising model and carefully examine how  the RBM extracts useful probabilistic  and physical information from Ising configurations. We find several indicators derived from the weight matrix that could characterize the Ising phase transition. We verify that the hidden encoding of a visible state tends to have an equal number of positive and negative units, whose sequence is randomly assigned during training and can be inferred by analyzing the weight matrix. We also explore the physical meaning  of visible energy and loss function (pseudo-likelihood) of the RBM and show that they could be harnessed to predict the critical point or estimate physical quantities such as  entropy. 
\end{abstract}


\maketitle

\section{Introduction}
The tremendous success of deep learning in multiple areas over the last decade has really revived  the interplay between physics and machine learning, in particular neural networks~\cite{carleo2019}.  On one hand, (statistical) physics ideas~\cite{bahri2020}, such as renormalization group (RG)~\cite{lin2017}, energy landscape~\cite{ballard2017}, free energy~\cite{zhang2018}, glassy dynamics~\cite{baity2018}, jamming~\cite{geiger2019}, Langevin dynamics~\cite{feng2021}, and field theory~\cite{roberts2022},  shed some light on the interpretation of deep learning and statistical inference in general~\cite{zdeborova2016}. On the other hand, machine learning and deep learning tools are harnessed to solved a wide range of physics problems, such as interaction potential construction~\cite{behler2007}, phase transition detection~\cite{carrasquilla2017}, structure encoding~\cite{bapst2020}, physical concepts discovery~\cite{iten2020}, and many others~\cite{bedolla2020,cichos2020}.  
At the very intersection of these two fields lies the restricted Boltzmann machine (RBM)~\cite{hinton2006}, which serves as a classical paradigm to investigate how an overarching perspective could benefit both sides.  

The RBM uses hidden-visible connections to encode  (high-order) correlations between visible units~\cite{smolensky1986}. Its  precursor--the (unrestricted) Boltzmann machine was inspired by  spin glasses~\cite{sherrington1975,ackley1985} and is often used in the inverse Ising problem to infer physical parameters~\cite{cocco2011,aurell2012,nguyen2017}.  The restriction of hidden-hidden and visible-visible connections in RBMs allows for more efficient training algorithms, and therefore leads to recent applications in  Monte Carlo simulation acceleration~\cite{huang2017}, quantum wavefunction representation~\cite{carleo2017,melko2019}, and polymer configuration generation~\cite{yu2019}.
Deep neural networks formed by stacks of RBMs have been  mapped onto the variational RG due to their conceptual similarity~\cite{mehta2014}. RBMs are also shown to be equivalent to tensor network states from quantum many-body physics~\cite{chen2018}.
As simple as it seems, energy-based models like the RBM could eventually become the building blocks of autonomous machine intelligence~\cite{lecun2022}.

Besides the above mentioned efforts, the RBM has also been applied extensively in the study of the minimal model for second-order phase transition--the Ising model. For the small systems under investigation, it was found that RBMs with an enough number of hidden units can encode the Boltzmann distribution, reconstruct thermal quantities, and generate new Ising configurations fairly well~\cite{torlai2016,morningstar2018,dangelo2020}. The visible $\to$ hidden $\to$ visible  $\cdots$ generating sequence of the RBM can be mapped onto  a RG flow in physical temperature (often towards the critical point)~\cite{iso2018,funai2020,koch2020}. But the mechanism and power   of the RBM to capture physics concepts and principles have not been fully explored. First, in what way is  the Boltzmann distribution of the Ising model learned by the RBM? Second, can the RBM learn and even quantitatively predict the phase transition without extra human knowledge? An affirmative answer to the second question is particularly appealing, because simple unsupervised learning methods such as principal component analysis (PCA) using configuration information alone do not provide quantitative prediction for the transition temperature~\cite{wang2016,wetzel2017} and supervised learning with neural networks requires human labeling of the phase type or temperature  of a given configuration~\cite{tanaka2017,kashiwa2019}.

In this article, we report a detailed numerical study on   RBM learning of the Ising model with a system size much larger than those used previously.
The purpose is to thoroughly dissect the various parts of the RBM and reveal how each part contributes to the learning of the Boltzmann distribution of the input Ising configurations. Such understanding allows us to extract several useful machine-learning estimators or predictors for physical quantities, such as entropy and phase transition temperature. Conversely, the analysis of a physical model helps us to obtain important insights about the meaning of RBM parameters and functions, such as weight matrix, visible energy and pseudo-likelihood. Below, we first introduce our Ising datasets,  the RBM and its training protocols in Sec.~\ref{sec:model}. We then report and discuss the results about model parameters, hidden layers, visible energy and pseudo-likelihood in 
Sec.~\ref{sec:result}. After the conclusion, more details about the Ising model and the RBM are provided in  Appendices.  Sample codes of the RBM are shared on the GitHub at \url{https://github.com/Jing-DS/isingrbm}.

\section{Models and Methods}
\label{sec:model}

\subsection{Dataset of Ising configurations generated by Monte Carlo simulations}
The Hamiltonian of the  Ising model with $N = L^d$ spins in a configuration ${\bf s}   = [s_1, s_2, \cdots, s_N]^T$ on a $d$-dimensional hypercubic lattice of linear dimension $L$ in the absence of magnetic field is
\begin{align}
{\mathcal H}( {\bf s} ) = -J \sum_{\langle i,j\rangle} s_i s_j
\end{align}
where the spin variable $s_i = \pm 1$ ($i=1,2,\cdots,N$),  the coupling parameter $J > 0$ (set to unity) favors ferromagnetic configurations (parallel spins) and the notation $\langle i,j  \rangle$ means to sum over nearest neighbors~\cite{cipra1987}. At a given temperature $T$, the configuration ${\bf s}$ drawn from the sample space of $2^N$ states follows the Boltzmann distribution
\begin{align}
p_T( {\bf s} ) = \frac{ e^{ - \frac{{\mathcal H}( {\bf s} ) }{k_BT} } }{Z_T}
\end{align}
where $Z_T = \sum\limits_{\bf s}  e^{ - \frac{{\mathcal H}( {\bf s} ) }{k_BT} } $ is the partition function. The Boltzmann constant $k_B$ is set to unity.

Using single-flip Monte Carlo simulations under periodic boundary conditions~\cite{newman:1999}, we generate Ising configurations for  two-dimensional ($2d$) systems ($d=2$) of $L = 64$ ($N = 4096$) at $n_T = 16$ temperatures $T=0.25, 0.5, 0.75, 1.0, \cdots, 4.0$ (in units of $J/k_B$) and for three-dimensional ($3d$)  systems $(d=3)$ of $L = 16$ ($N = 4096$) at $n_T = 20$ temperatures $T =2.5,2.75$,$3.0,3.25$,$3.5,3.75,4.0$, $4.25,4.3,4.4,4.5$,$4.6,4.7,4.75,5.0,5.25$,$5.5,5.75,6.0, 6.25$. After fully equilibrated, $M=50000$  configurations at each $T$ are collected into a dataset $D_T$ for that $T$. For $2d$ systems, we also use a dataset $D_{\cup T}$ consisting of $50000$ configurations per temperature from {\em all} $T$'s.

Analytical results about thermal quantities of the $2d$ Ising model, such as internal energy $\langle E \rangle$, (physical) entropy $S$, heat capacity $C_V$ and magnetization $\langle m\rangle$, are well known~\cite{kramers:1941,onsager1944,yang:1952,plischoke:1994}. Numerical simulation methods and results about the $3d$ Ising model have also been reported~\cite{landau2021}. Thermodynamic definitions and relations used in this work are summarized in Appendix~\ref{ap:ising}.

\subsection{Restricted Boltzmann Machine (RBM)}
\begin{figure}
\includegraphics[width=0.45\textwidth]{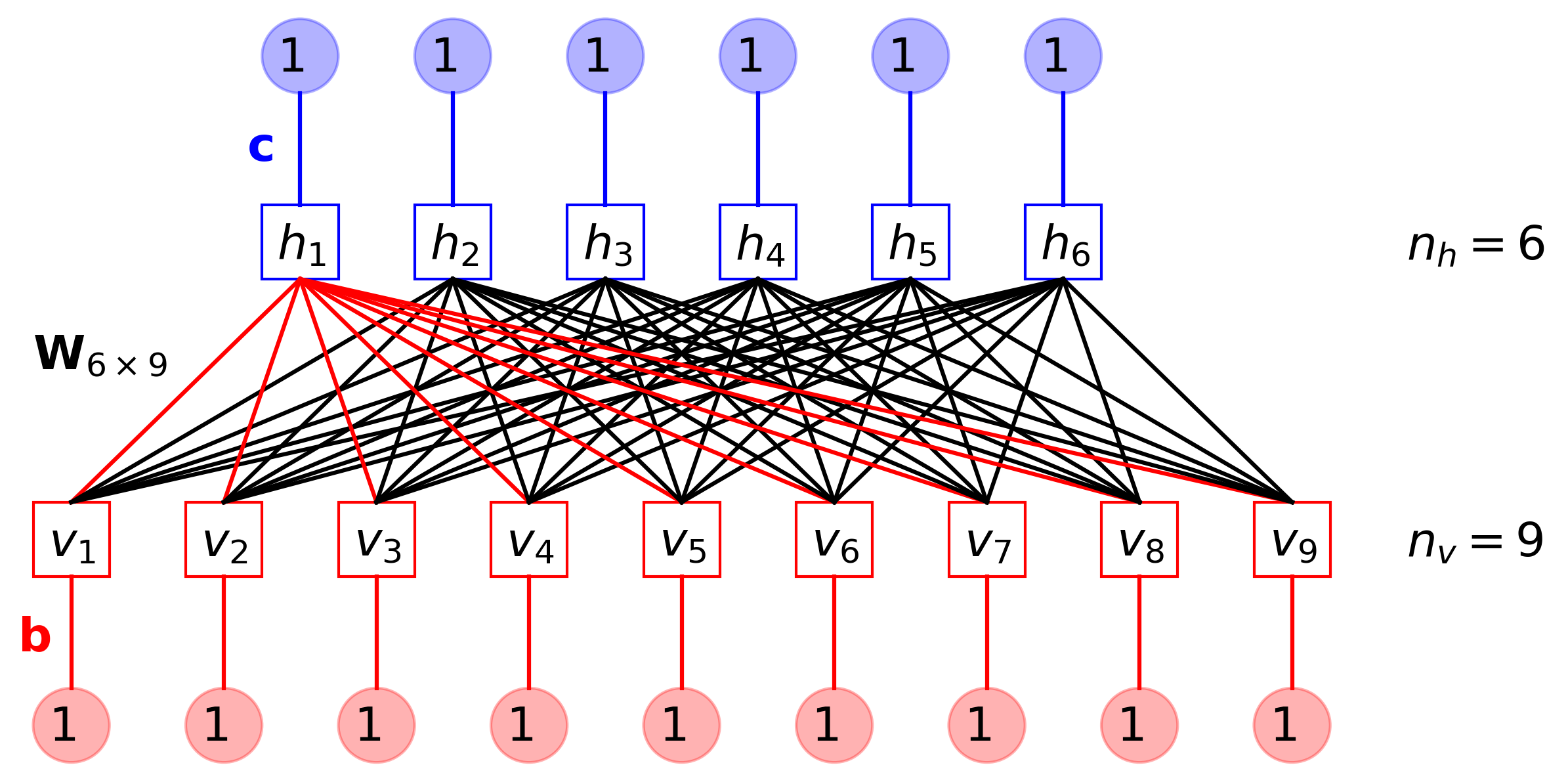}
\caption{A restricted Boltzmann machine (RBM) with $n_h=6$ hidden units and $n_v = 9$ visible units. Model parameters ${\boldsymbol\theta} = \{ {\bf W}, {\bf b}, {\bf c} \}$ are represented by connections. A filter ${\bf w}_{1}^T$  from visible units to the first hidden unit is highlighted by red (light color) connections.}
\label{fig:rmb}
\end{figure}

The restricted Boltzmann machine (RBM) is a two-layer energy-based model with $n_h$ hidden units (or neurons)  $h_i=\pm 1$ ($i=1,2,\cdots, n_h$) in the hidden layer, whose state vector is ${\bf h} = [h_1, h_2, \cdots, h_{n_h}]^T$,  and $n_v$ visible units  $v_j=\pm 1$ ($j=1,2,\cdots,n_v$) in the visible layer, whose state vector is ${\bf v} = [v_1, v_2, \cdots, v_{n_v}]^T$  (Fig.~\ref{fig:rmb})~\cite{fischer2012}. In this work, the visible layer is just the Ising configuration vector, i.e. ${\bf v} =  {\bf s}$, with  $n_v = N$. We choose binary unit $\{-1, +1 \}$ (instead of $\{0 ,1 \}$) to better align with the  definition of Ising spin variable $s_i$.

The  total energy  $E_{\boldsymbol{\theta}}({\bf v}, {\bf h})$ of the RBM  is defined as
\begin{equation}
\begin{aligned}
E_{\boldsymbol{\theta}}({\bf v}, {\bf h}) &= - {\bf b}^T {\bf v} - {\bf c}^T {\bf h} - {\bf h}^T{\bf W} {\bf v} \\
&=   -   \sum\limits_{j=1}^{n_v} b_j  v_j - \sum\limits_{i=1}^{n_h} c_i h_i  -\sum\limits_{i=1}^{n_h}  \sum\limits_{j=1}^{n_v} W_{ij} h_i v_j
\end{aligned}
\end{equation}
where ${\bf b} = [b_1, b_2, \cdots, b_{n_v}]^T$ is the visible bias, ${\bf c} = [c_1, c_2, \cdots, c_{n_h}]^T$ is the hidden bias and
\begin{equation}
{\bf W}_{n_h \times n_v} = 
\begin{bmatrix}
-{\bf w}_1^T-\\
-{\bf w}_2^T-\\
\vdots\\
-{\bf w}_{n_h}^T-\\
\end{bmatrix}
=
\begin{bmatrix}
| & | & & |\\
{\bf w}_{:, 1} & {\bf w}_{:, 2}  & \cdots & {\bf w}_{:, n_v} \\
| & | & & |
\end{bmatrix}
\end{equation}
 is the interaction weight matrix between visible and hidden units. Under this notation, each row vector ${\bf w}_{i}^T$ (of dimension $n_v$) is a {\em filter} mapping from the visible state ${\bf v}$ to a hidden unit $i$ and each column vector  ${\bf w}_{:, j}$ (of dimension $n_h$) is an {\em inverse filter} mapping from  the hidden state ${\bf h}$ to a visible unit $j$.
All parameters are collectively written as ${\boldsymbol\theta} = \{ {\bf W}, {\bf b}, {\bf c} \}$.``Restricted'' refers to the lack of interaction between   hidden units or between   visible units.

The joint distribution for an overall state $({\bf v}, {\bf h})$ is
\begin{equation}
p_{\boldsymbol \theta}({\bf v}, {\bf h}) = \frac{e^{- E_{\boldsymbol \theta}({\bf v}, {\bf h})}}{Z_{\boldsymbol \theta}}
\end{equation}
where the partition function of the RBM 
\begin{equation}
Z_{\boldsymbol \theta} = \sum_{\bf v} \sum_{\bf h} e^{- E_{\boldsymbol \theta}({\bf v}, {\bf h})}.
\end{equation}
The learned {\em model distribution} for visible state ${\bf v}$ is from   marginalization of  $p_{\boldsymbol \theta}({\bf v}, {\bf h})$, 
\begin{equation}
p_{\boldsymbol \theta}({\bf v})  = \sum_{\bf h} p_{\boldsymbol \theta}({\bf v}, {\bf h})  =  \frac{1}{Z_{\boldsymbol \theta}}  e^{- {\mathcal E}_{\boldsymbol \theta}({\bf v})},
\end{equation}
where the {\em visible energy}--an effective energy for visible state ${\bf v}$ (often termed as ``free energy'' in machine learning literature),
 \begin{equation}
\begin{aligned}
 {\mathcal E}_{\boldsymbol \theta}({\bf v}) 
&= -{\bf b}^T {\bf v} - \sum\limits_{i=1}^{n_h}  \ln \left( e^{   - {\bf w}_i^T {\bf v} - c_i  }  +  e^{    {\bf w}_i^T {\bf v} + c_i     } \right)
\end{aligned}
\end{equation}
 is  defined according to $e^{- {\mathcal E}_{\boldsymbol \theta}({\bf v}) } = \sum\limits_{\bf h} e^{- E_{\boldsymbol \theta}({\bf v}, {\bf h})}$ such that
$Z_{\boldsymbol \theta} = \sum\limits_{\bf v}  e^{- {\mathcal E}_{\boldsymbol \theta}({\bf v})}.$
See Appendix~\ref{ap:rbm} for a detailed derivation.

The conditional distributions to generate ${\bf h}$ from ${\bf v}$,  $p_{\boldsymbol \theta}({\bf h}|{\bf v})$, and to generate ${\bf v}$ from ${\bf h}$,  $p_{\boldsymbol \theta}({\bf v}|{\bf h})$, satisfying $  p_{\boldsymbol \theta}({\bf v},{\bf h}) = p_{\boldsymbol \theta}({\bf h}|{\bf v})  p_{\boldsymbol \theta}({\bf v}) =  p_{\boldsymbol \theta}({\bf v}|{\bf h})p_{\boldsymbol \theta}({\bf h})$, can be written as products  
\begin{equation}
\begin{aligned}
p_{\boldsymbol \theta}({\bf h}|{\bf v})  &= \prod\limits_{i=1}^{n_h} p_{\boldsymbol \theta}(h_i | {\bf v}) \\
 p_{\boldsymbol \theta}({\bf v}|{\bf h}) &= \prod\limits_{j=1}^{n_v} p_{\boldsymbol \theta}(v_j | {\bf h})
 \end{aligned}
\end{equation}
because $h_i$ are independent from each other (at fixed ${\bf v}$) and $v_j$ are independent from each other (at fixed ${\bf h}$).
It can be shown that
\begin{equation}
\label{eq:cond_prob}
\begin{aligned}
p_{\boldsymbol \theta}(h_i = 1 | {\bf v}) &= \sigma\left( 2(c_i +  {\bf w}_{i}^T {\bf v} ) \right) \\
p_{\boldsymbol \theta}(h_i = -1 | {\bf v}) & =  1- \sigma\left( 2( c_i +  {\bf w}_{i}^T {\bf v} ) \right) \\
p_{\boldsymbol \theta}(v_j = 1 | {\bf h}) &= \sigma\left( 2 ( b_j +  {\bf h}^T {\bf w}_{:,j} ) \right) \\
 p_{\boldsymbol \theta}(v_j = -1 | {\bf h}) &= 1- \sigma\left( 2( b_j +  {\bf h}^T {\bf w}_{:,j} ) \right)
\end{aligned}
\end{equation}
where the sigmoid function $\sigma(z) = \frac{1}{1+e^{-z}}$ (Appendix~\ref{ap:rbm}).

\subsection{Loss function and training of RBMs}
Given the dataset $D = [ {\bf v}_1, {\bf v}_2, \cdots, {\bf v}_M ]^T$ of $M$ samples generated independently from the identical {\em data distribution} $p_D({\bf v})$ (${\bf v} \stackrel{\text{i.i.d.}}{\sim} p_D({\bf v})$), the  goal of RBM learning is to find a model distribution $p_{\boldsymbol \theta}({\bf v})$ that approximates $p_D({\bf v})$. In the context of this work, the data samples ${\bf v}$'s are Ising configurations and the data distribution $p_D({\bf v})$ is or is related to the Ising Boltzmann distribution $p_T( {\bf s} )$. 

Based on maximum likelihood estimation, the optimal parameters ${\boldsymbol \theta}^* =  \arg \min\limits_{\boldsymbol \theta}  {\mathcal L}({\boldsymbol \theta}) $ can be found by minimize the negative log likelihood
\begin{equation}
\label{eq:L}
 {\mathcal L}({\boldsymbol \theta}) = \langle -\ln p_{\boldsymbol \theta}({\bf v}) \rangle_{{\bf v}\sim p_D} = \langle  {\mathcal E}_{\boldsymbol \theta}({\bf v})  \rangle_{{\bf v}\sim p_D}  + \ln Z_{\boldsymbol \theta}
\end{equation}
which serves as the {\em loss function} of RBM learning. Note that the partition function $Z_{\boldsymbol \theta}$ only depends on the model but not on data. Since the calculation of  $Z_{\boldsymbol \theta}$ involves summation over all possible   $({\bf v}, {\bf h})$ states, which is not feasible, $ {\mathcal L}({\boldsymbol \theta})$ can not be evaluated exactly, except for very small systems~\cite{oh2020}.  Approximations have to be made, for example, by  mean-field calculations~\cite{huang2015}.  An interesting feature of the RBM is that, although the actual loss function $  {\mathcal L}({\boldsymbol \theta})$ is not accessible, its gradient 
\begin{equation}
 \nabla_{\boldsymbol \theta}   {\mathcal L}({\boldsymbol \theta}) = \langle  \nabla_{\boldsymbol \theta}  {\mathcal E}_{\boldsymbol \theta}({\bf v})  \rangle_{{\bf v}\sim p_D}  - \langle  \nabla_{\boldsymbol \theta}  {\mathcal E}_{\boldsymbol \theta}({\bf v})  \rangle_{{\bf v}\sim p_\theta}
\end{equation}
can be sampled, which enables a gradient descent learning algorithm.  From step $t$ to step $t+1$,  model parameters are updated with learning rate $\eta$ as
\begin{equation}
{\boldsymbol \theta}_{t+1} =  {\boldsymbol \theta}_t - \eta  \nabla_{\boldsymbol \theta}   {\mathcal L}({\boldsymbol \theta}_t).
\end{equation}

To evaluate the loss function, we use its approximate -- the pseudo-(negative log)likelihood~\cite{besag1975}
\begin{equation}
\widetilde{\mathcal L}({\boldsymbol \theta}) = \left \langle-  \sum_{i=1}^{n_v} \ln p_{\boldsymbol \theta} (v_i | v_{j\ne i}) \right \rangle_{{\bf v}\sim p_D}
\approx  {\mathcal L}({\boldsymbol \theta})
\end{equation}
where the notation 
\begin{equation}
\begin{aligned}
p_{\boldsymbol \theta} (v_i | v_{j \ne i}) &= p_{\boldsymbol \theta} (v_i | v_j ~ {\rm for} ~j \ne i) \\
& =  \frac{ e^{- {\mathcal E}_{\boldsymbol \theta}({\bf v})}   }{  e^{- {\mathcal E}_{\boldsymbol \theta}({\bf v})}  + e^{- {\mathcal E}_{\boldsymbol \theta}( [v_1,  \cdots, -v_i, \cdots, v_{n_v}])} }
\end{aligned}
\end{equation}
is the conditional probability for component $v_i$ given that all the other components  $v_j$ $(j \ne i)$ are fixed.  
Practically, to avoid the time-consuming sum over all visible units $\sum\limits_{i=1}^{n_v} $, it is suggested to randomly sample one $i_0 \in \{ 1, 2, \cdots, n_v \}$ and estimate that
\begin{equation}
\widetilde{\mathcal L}({\boldsymbol \theta}) \approx \left \langle-   n_v \ln p_{\boldsymbol \theta} (v_{i_0} | v_{j \ne i_0}) \right \rangle_{{\bf v}\sim p_D},
\end{equation}
if all the visible units are on average translation-invariant~\cite{lisa2018}. 
To monitor the reconstruction error, we also calculate the cross entropy ${\rm CE}$ between the initial configuration ${\bf v}$ and the conditional probability $p_{\boldsymbol \theta}( {\bf v}'  | {\bf h})$ for reconstruction ${\bf v} \stackrel{p_{\boldsymbol \theta}( {\bf h}  | {\bf v})}{\longrightarrow} {\bf h} \stackrel{p_{\boldsymbol \theta}( {\bf v}'  | {\bf h})}{\longrightarrow} {\bf v}'$ (See Appendix~\ref{ap:rbm_train} for definition).

For both $2d$ and $3d$ Ising systems, we first train single temperature RBMs ($T$-RBM). $M=50000$ Ising configurations at each $T$  forming a dataset $D_T$ are used to train one model such that there are $n_T$ $T$-RBMs in total. While $n_v = N$, we try various number of hidden units with $n_h = 400, 900, 1600, 2500$ in $2d$ and $n_h = 400, 900, 1600$ in $3d$. For $2d$ systems, we also train an all temperature RBM ($\cup T$-RBM) for which $50000$ Ising configurations per temperature are drawn to compose a dataset $D_{\cup T}$ of $M = 50000 n_T = 8\times 10^5$ samples.  The number of hidden units  for this $\cup T$-RBM is $n_h = 400, 900, 1600.$  Weight matrix ${\bf W}$ are initialized with Glorot normal
initialization~\cite{glorot2010} (${\bf b}$ and ${\bf c}$ are initialized as zero). Parameters are optimized with the stochastic gradient descent algorithm of  learning rate $\eta = 1.0\times 10^{-4}$ and batch size 128. The negative phase (model term) of the gradient $\langle  \nabla_{\boldsymbol \theta}  {\mathcal E}_{\boldsymbol \theta}({\bf v})  \rangle_{{\bf v}\sim p_\theta}$ is calculated using  CD-k Gibbs sampling  with $k=5$. We stop the training  until $\widetilde{\mathcal L}$ and CE converge, typically at 100-2000   epochs (see Supplemental Material).
Three Nvidia GPU cards (GeForce RTX 3090 and 2070) are used to train the model, which takes about two mins per epoch for a $M=50000$ dataset.

\section{Results and Discussion}
\label{sec:result}
In this section, we investigate how the RBM uses its weight matrix ${\bf W}$ and hidden layer ${\bf h}$ to encode the Boltzmann distributed states of the Ising model, and what physical information can be extracted from machine learning concepts such as visible energy and loss function.

\subsection{Filters and inverse filters}
\begin{figure}
\includegraphics[width=0.45\textwidth]{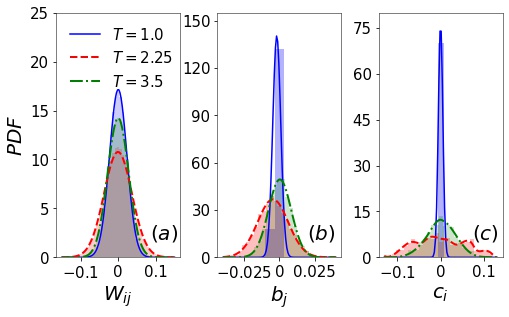}
\caption{  Probability density function (PDF) of the distribution of (a) $W_{ij}$,  (b) $b_j$, and (c)  $c_i$ of $T$-RBMs with $n_h=400$ hidden units at temperatures below, close to and above $T_c$.}
\label{fig:ising2d_W_b_c}
\end{figure}

It can be verified that the trained weight matrix elements $W_{ij}$ of a $T$-RBM follows a Gaussian distribution of zero mean with largest variance at $T \sim T_c$ (Fig.~\ref{fig:ising2d_W_b_c}a)  The high temperature distribution here is different from the uniform distribution observed in Ref.~\cite{torlai2016}. According to Eq.~(\ref{eq:cond_prob}), the biases $c_i$ and $b_j$ can be associated with the activation threshold of a hidden unit and a visible unit, respectively. For example,  whether a hidden unit is activated ($h_i = +1$) or anti-activated  ($h_i = -1$) depends on whether the incoming signal ${\bf w}^T_i {\bf v}$ from all visible units exceeds the threshold $-c_i$. The values of $c_i$ (and $b_j$) are all close to zero and  are  often negligible in comparison  with the total incoming signal ${\bf w}^T_i {\bf v}$ (and ${\bf h}^T {\bf w}_{:,j}$) (see Supplemental Material for the results of constrained RBMs where all biases are  set to zero). The distribution of  $c_i$ and $b_j$ should in principle be symmetric about zero (Fig.~\ref{fig:ising2d_W_b_c}b-c).  A non-zero mean can be caused by an unbalanced dataset with unequal number of $m>0$ and $m<0$ Ising configurations.  The corresponding filter or inverse filter sum may also be distributed with a non-zero mean in order to compensate the asymmetric bias as will be shown next.

\begin{figure}
\includegraphics[width=0.45\textwidth]{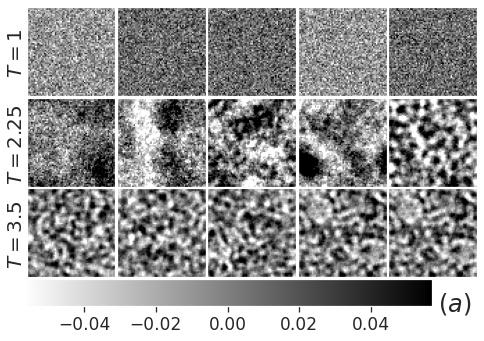}
\includegraphics[width=0.45\textwidth]{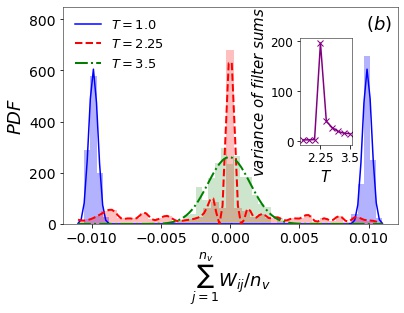}
\includegraphics[width=0.45\textwidth]{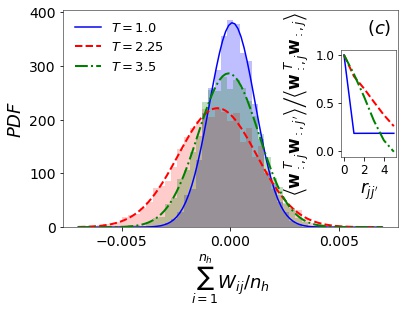}
\caption{$T$-RBMs  with  $n_h=400$  for the $2d$ Ising model at temperature $T=1.0$, $2.25$, and $3.5$. (a) Five sample filters ${\bf w}^T_i$  at each temperature. The colorbar range is set to be within about two standard deviations of the distribution. (b) PDF of the distribution of the  $n_h = 400$ filter sums (normalized by $n_v$). Inset: variance $ \left \langle \left| \sum\limits_{j=1}^{n_v}  W_{ij}  \right|^2 \right\rangle - \left \langle \left| \sum\limits_{j=1}^{n_v}  W_{ij}   \right| \right\rangle^2$ of the filter sum as a funciton of temperature. (c) PDF of the distribution of the  $n_v = 4096$ inverse filter sums (normalized by $n_h$). Inset: correlation between a pair of inverse filters ${\bf w}_{:,j}$ and ${\bf w}_{:,j'}$ (normalized by auto-correlation) as a function of spin-spin distance $r_{jj'}$. } 
\label{fig:ising2d_W_sample}
\end{figure}

Since ${\bf v} = {\bf s}$ is an Ising configuration with $\pm 1$ units in our problem, ${\bf w}^T_i {\bf v}$  will be more positive (or negative) if  the components of ${\bf w}^T_i$ better match (or anti-match) the signs of spin variables. In this sense, we can think of  ${\bf w}^T_i$ as a filter extracting certain patterns in Ising configurations. Knowing the representative spin configurations of the Ising model below, close to and above the critical temperature $T_c$, we expect that ${\bf w}^T_i$ ($i=1,2,\cdots, n_h$) wrapped into a $L^d$ arrangement exhibits similar features.   In Fig.~\ref{fig:ising2d_W_sample}a, we show sample filters of $T$-RBMs with $n_h = 400$ trained for the $2d$ Ising model at three temperatures $T=1.0, 2.25$ and $3.5$ (see Supplemental Material for more examples of filters). At low $T$, the components of ${\bf w}^T_i$ tend to be mostly positive (or negative) matching the spin up (or spin down) configurations in the ferromagnetic phase. At high $T$, filters ${\bf w}^T_i$ possess strip domains consisting of roughly  equal number of well-mixed positive and negative components, like Ising configurations during spinodal decomposition. Close to $T_c$, ${\bf w}^T_i$  patterns vary dramatically from each other,  in accord with the large critical fluctuation. In particular, some even exhibit hierarchical clusters of various sizes. The element sum of the filter -- {\em filter sum} ${\rm sum}({\bf w}_i^T ) = \sum\limits_{j=1}^{n_v}  W_{ij} $, plays the similar role as the magnetization $m$. The distribution of all the $n_h$ filter sums at each $T$ changes with increasing temperature as the Ising magnetization changes, from bimodal to unimodal with largest variance at $T_c$ (Fig.~\ref{fig:ising2d_W_sample}b). This suggests that the peak of  the variance $ \left \langle \left| \sum\limits_{j=1}^{n_v}  W_{ij} \right|^2 \right\rangle - \left \langle \left| \sum\limits_{j=1}^{n_v}  W_{ij} \right| \right\rangle^2$ as a function of temperature coincides with the Ising phase transition (inset of Fig.~\ref{fig:ising2d_W_sample}b). More detailed results about $2d$ and $3d$ Ising model are in Supplemental Material.

\begin{figure}
\includegraphics[width=0.4\textwidth]{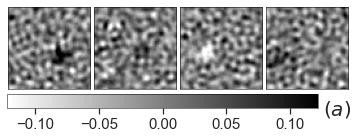}
\includegraphics[width=0.45\textwidth]{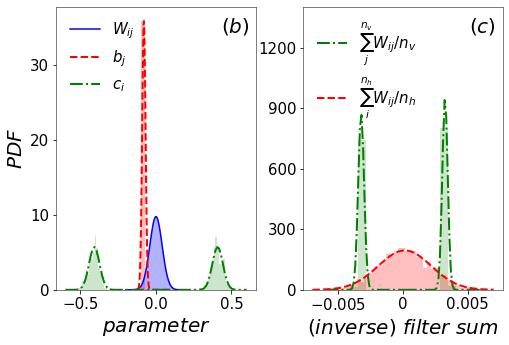}
\caption{The $\cup T$-RBM  with  $n_h=400$  for the $2d$ Ising model. (a) Four sample filters ${\bf w}^T_i$. (b) PDF of the distribution of $W_{ij}$,  $b_j$, and $c_i$. (c) PDF of the distribution of the  $n_h = 400$ filter sums and the $n_v = 4096$ inverse filter sums. }
\label{fig:ising2d_all_W_sample}
\end{figure}

When a  hidden layer ${\bf h}$ is provided, the RBM reconstructs the visible layer ${\bf v}$ by applying the $n_v$ inverse filters ${\bf w}_{:, j}$ ($j=1,2,\cdots,n_v$) on ${\bf h}$. The distribution of the {\em inverse filter sum} ${\rm sum}({\bf w}_{:,j}) = \sum\limits_{i=1}^{n_h}  W_{ij} $ is Gaussian with a mean close to zero (Fig.~\ref{fig:ising2d_W_sample}c), where a large deviation from zero mean is accompanied by a non-zero average bias $ \sum\limits_j b_j /n_v$ as mentioned above (Fig.~\ref{fig:ising2d_W_b_c}b). We find that this is a result of the unbalanced dataset which has $\sim 60\%$ $m<0$ Ising configurations. Because the activation probability of a visible unit $v_j$ is determined by ${\bf w}_{:, j}$, the correlation between visible units (Ising spins) is reflected in the correlation between inverse filters. This is equivalent to the analysis of the $n_v\times n_v$ matrix ${\bf W}^T {\bf W}$ as in Ref.~\cite{iso2018}, whose entries are inner product ${\bf w}_{:,j}^T {\bf w}_{:,j'}$ of inverse filters.  We can therefore locate the Ising phase transition by identifying the temperature with the strongest correlation among ${\bf w}_{:, j}$'s, e.g. the peak of ${\bf w}_{:, j}^T {\bf w}_{:, j'}$ at a given distance $r_{jj'}$ (inset of Fig.~\ref{fig:ising2d_W_sample}c). See Supplemental Material for results in $2d$ and $3d$.

In contrast, the filters of the $\cup T$-RBM trained from $2d$ Ising configurations at all temperatures have background patterns like the high temperature $T$-RBM (in the paramagnetic phase). A clear difference is that most $\cup T$-RBM filters have one large domain of positive or negative elements (Fig.~\ref{fig:ising2d_all_W_sample}a), similar as the receptive field in a deep neural network~\cite{mehta2014}. This domain randomly covers an area of the visual field of the $L\times L$ Ising configuration (see Supplemental Material for all the $n_h$ filters).  The existence of such domains in the filter causes the filter sum and the corresponding bias $c_i$ to be positive or negative with a bimodal distribution (Fig.~\ref{fig:ising2d_all_W_sample}b-c). The inverse filter sum and its corresponding bias $b_j$ still has a Gaussian distribution, although the unbalanced dataset  shifts the mean of $b_j$ away from zero.

\subsection{Hidden layer}
\begin{figure}
\includegraphics[width=0.45\textwidth]{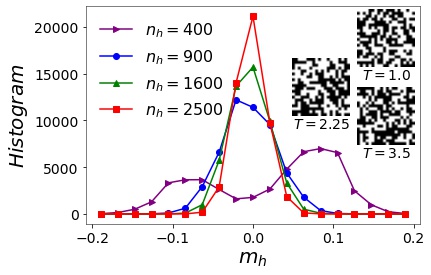}
\caption{Histogram of $m_h$ obtained from the hidden encodings of  $M=50000$  $2d$ Ising configurations at $T=2.25$ using $T$-RBMs with various $n_h$. Inset: examples of the hidden layer of $T$-RBMs with $n_h=400$ wrapped into a $20 \times 20$ matrix at three temperatures, where $+1$/$-1$ units are represented by black/white pixels.}
\label{fig:ising2d_var_m_h}
\end{figure}

Whether a hidden unit uses $+1$ or $-1$ to encode a pattern of the visible layer ${\bf v}$ is randomly assigned during training. In the former case, the filter ${\bf w}^T_i$ matches the pattern (${\bf w}^T_i {\bf v}$ is positive);   in the latter case, the filter anti-matches the pattern (${\bf w}^T_i {\bf v}$ is negative).  For a  visible layer ${\bf v}$ of  magnetization $m$, the  sign of $ {\bf w}_i^T {\bf v}$ and the encoding $h_i$ is largely determined by the sign of ${\rm sum}({\bf w}_i^T )$ (Table~\ref{table:hidden}). Since the distribution of ${\rm sum}({\bf w}_i^T )$ is  symmetric about zero, the hidden layer of a $T$-RBM   roughly consists of an equal number of $+1$  and $-1$ units -- the ``magnetization'' $m_h = \frac{1}{n_h}\sum\limits_{i=1}^{n_h} h_i$ of the hidden layer is always close to zero and its average $\langle m_h \rangle \approx 0$.  The histogram of $m_h$ for all hidden encodings of   visible states is expected to be symmetric about zero (Fig.\ref{fig:ising2d_var_m_h}). We find that for the smallest $n_h$ the histogram of $m_h$ at temperatures close to $T_c$ is bimodal due to the relatively large randomness of small hidden layers. As more hidden units are added, the two peaks merge into one and the distribution of $m_h$ becomes narrower. This suggests that larger hidden layer tends to have smaller deviation from $m_h=0$. 
\begin{table}[H]
\setlength\extrarowheight{4pt}
\caption{When ${\rm sum}({\bf w}_i^T )>0$, a visible layer pattern ${\bf v}$ with magnetization $m>0$ (or $m<0$) is more likely to be encoded by a hidden unit $h_i = +1$ (or $h_i = -1$). When ${\rm sum}({\bf w}_i^T )<0$, the encoding is opposite.}
\begin{center}
\begin{tabular}{ccc}
\hline 
  & ${\rm sum}({\bf w}_i^T )>0$&  ${\rm sum}({\bf w}_i^T )<0$  \\
\hline 
$m > 0$  & $h_i = +1$ & $h_i = -1$\\
$m < 0$  & $h_i = -1$ & $h_i = +1$\\
\hline
\end{tabular}
\end{center}
\label{table:hidden}
\end{table}

The order of the $h_i=\pm 1$ sequence in each hidden encoding ${\bf h}$ is arbitrary but relatively fixed once the $T$-RBM is trained. Permutation of hidden units together with their corresponding filters (swap rows of the matrix ${\bf W}$) results in an equivalent $T$-RBM. Examples of hidden layers of  $T$-RBMs with $n_h=400$ at different temperatures are shown in the inset of  Fig.\ref{fig:ising2d_var_m_h}, where the vector ${\bf h}$ is wrapped into a $20\times 20$ arrangement. Note that there is actually no spatial relationships between different hidden units and any apparent pattern in this $2d$ illustration is an artifact of the wrapping protocol.

\begin{figure}
\includegraphics[width=0.4\textwidth]{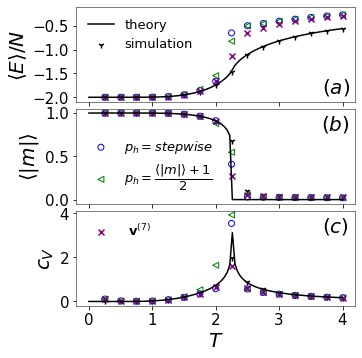}
\caption{(a) Internal energy, (b) magnetization, and (c) specific heat of $2d$ Ising states reconstructed by $T$-RBMs ($n_h=400$) with the hidden layer ${\bf h}^{(0)}$ initiated according to  $p_h = ( \langle |m| \rangle +1)/2$ or $p_h = 1.0 (T\le2.0), 0.5 (T \ge 2.5), 0.75 (2.0<T<2.5)$ (stepwise). Reconstruction  by a seven-step Markov chain from random  ${\bf h}^{(0)}$ is compared (${\bf v}^{(7)}$). Analytical and Monte Carlo simulation results are also shown.  } 
\label{fig:ising2d_generate}
\end{figure}

\begin{figure*}
\includegraphics[width=1\textwidth]{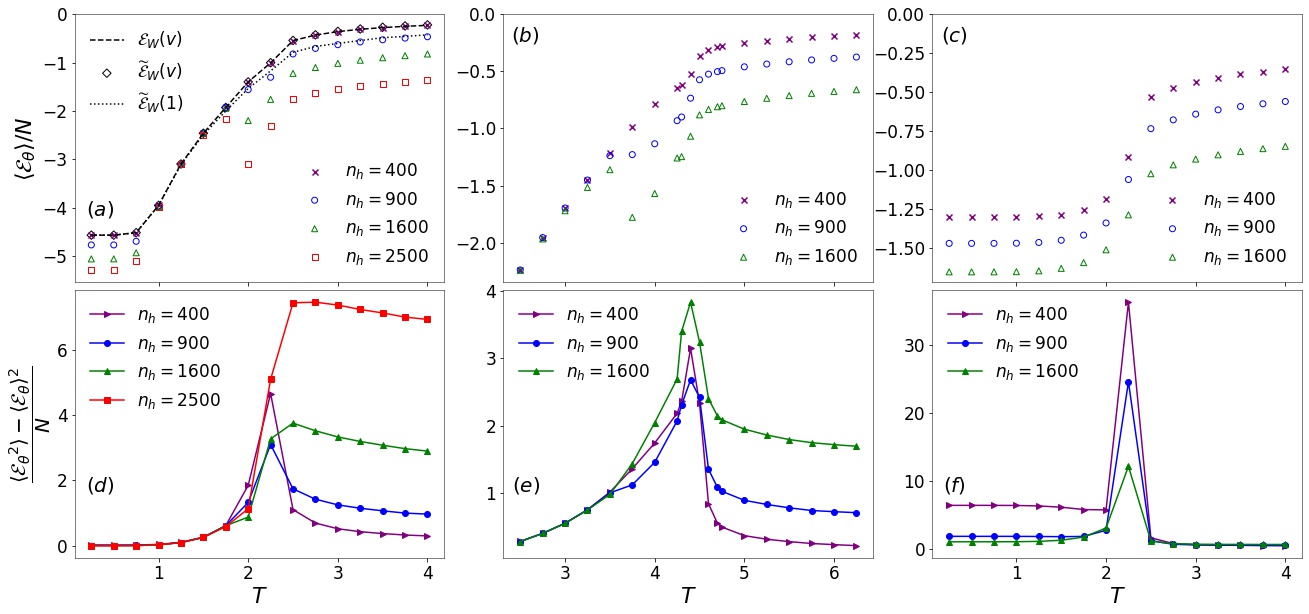}
\caption{Mean and variance of visible energy ${\mathcal E}_{\boldsymbol \theta}$ as a function of temperature for  $2d$ (a,c,d,f) and $3d$ (b,e) Ising models captured by $T$-RBMs (a,b,d,e) and the $\cup T$-RBM (c,f) of various hidden neurons $n_h$. Three approximate forms of visible energy for $n_h=400$ $T$-RBMs are shown in (a).}
\label{fig:combined_rbm_E}
\end{figure*}

As a generative model, a $T$-RBM can be used to produce more Boltzmann distributed Ising configurations. Starting from a random hidden state ${\bf h}^{(0)}$, this is often fulfilled by  a sequence of Markov chain moves ${\bf h}^{(0)} \to {\bf v}^{(0)} \to {\bf h}^{(1)} \to {\bf v}^{(1)} \to \cdots $ until steady state is achieved. Based on above mentioned observations, we can design an algorithm to initialize ${\bf h}^{(0)}$ that  better captures the hidden encoding of visible states (equilibrium Ising configurations), thus enables faster convergence of the Markov chain. After choosing a low temperature $T_L$ and a high temperature $T_H$, we generate the hidden layer as follows:
\begin{itemize}
    \item At low $T \le T_L < T_c$, if ${\rm sum}({\bf w}_i^T )>0$, $h_i = + 1$; if ${\rm sum}({\bf w}_i^T )<0$, $h_i = - 1$. This will be an encoding of a $m>0$ ferromagnetic configuration.  To encode of a $m<0$ ferromagnetic configuration, just flip the sign of $h_i$.
    \item At high $T \ge T_H > T_c$, randomly assign  $h_i = + 1$ or $-1$ with equal probability. This will be an encoding of a paramagnetic configuration with $m \approx 0$.
    \item At intermediate  $T_L < T < T_H$, to encode a  $m>0$ Ising configuration, if ${\rm sum}({\bf w}_i^T )>0$, assign $h_i = + 1$ with probability $p_h \in (0.5,1.0)$ and $h_i = - 1$ with probability $1-p_h$;  if ${\rm sum}({\bf w}_i^T )<0$, assign $h_i = - 1$ with probability $p_h \in (0.5,1.0)$ and $h_i = + 1$ with probability $1-p_h$. $p_h$ is a predetermined parameter and the above two algorithms  are just the special cases with $p_h = 1.0$ ($T \le T_L$) and $p_h = 0.5$ ($T \ge T_H$), respectively.  In practice, one may approximately use  $p_h = ( \langle |m| \rangle +1)/2$ or use linear interpolation within $T_L < T < T_H$, $p_h = 0.5 + 0.5(T-T_L)/(T_H - T_L)$.
\end{itemize}
Below we compare the (one-step) reconstructed thermal quantities using two different initial hidden encodings with results from a conventional multi-step Markov chain (Fig.~\ref{fig:ising2d_generate}).  The hidden encoding methods proposed here are quite reliable at low and high $T$, but less accurate at $T$ close to  $T_c$.

\begin{figure*}
\centering
\includegraphics[width=1\textwidth]{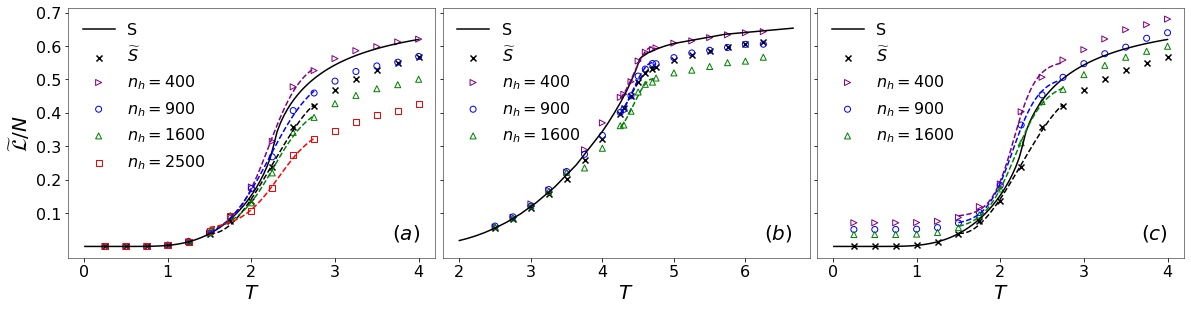}
\caption{Pseudo-likelihood $\widetilde{\mathcal L}$ per spin of $T$-RBMs (a-b) and of the $\cup T$-RBM (c) with different number $n_h$ of hidden units for the $2d$ (a,c) and $3d$ (b) Ising model in comparison with entropy $S$ and pseudo-entropy $\widetilde{S}$ per spin. Dashed lines are polynomial fittings around $T_c$.}
\label{fig:ising_L}
\end{figure*}

\subsection{Visible energy}

When a $T$-RBM for temperature  $T$ is trained,  we expect that $p_{\boldsymbol \theta}({\bf v}) \approx p_D({\bf v}) \approx   p_T({\bf s})$ -- the Boltzmann distribution at that $T$.
Although formally related to the physical energy  in the Boltzmann factor (with temperature absorbed), the visible energy $ {\mathcal E}_{\boldsymbol \theta}({\bf v}) $ of a RBM should be really considered as the  negative log (relative) probability of a visible state ${\bf v}$. For single temperature $T$-RBMs, the mean visible energy $ \langle {\mathcal E}_{\boldsymbol \theta}({\bf v}) \rangle $ increases monotonically with temperature (except for the largest $n_h$, which might be due to overfitting) (Fig.~\ref{fig:combined_rbm_E}a-b). The value of $ \langle {\mathcal E}_{\boldsymbol \theta}({\bf v}) \rangle $ and its trend, however, cannot be used to identify the physical phase transition. In fact,  $ {\mathcal E}_{\boldsymbol \theta}({\bf v}) $  can differ from the reduced Hamiltonian $\frac{\mathcal H({\bf s})}{k_BT}$ by an arbitrary (temperature-dependent) constant while still maintaining the Boltzmann distribution  $p_{\boldsymbol \theta}({\bf v}) \approx p_T({\bf s})$ (if the partition function $Z_{\boldsymbol \theta}$ is calibrated accordingly).

The trend of $ \langle {\mathcal E}_{\boldsymbol \theta}({\bf v}) \rangle $ for $T$-RBMs can be understood by considering following approximate forms. First, due to the symmetry of $+1$ and $-1$, the biases $b_j$ and $c_i$ are all close to zero. A constrained $T$-RBM with zero bias has a visible energy 
\begin{equation}
\label{eq:Ew}
\begin{aligned}
 {\mathcal E}_{\boldsymbol W}({\bf v}) 
= - \sum\limits_{i=1}^{n_h}  \ln \left( e^{   - {\bf w}_i^T {\bf v}}  +  e^{    {\bf w}_i^T {\bf v}} \right)
\end{aligned}
\end{equation}
that approximates the visible energy of the full $T$-RBM, i.e. $ {\mathcal E}_{\boldsymbol \theta}({\bf v})   \approx {\mathcal E}_{\boldsymbol W}({\bf v}) $.
Next, unless ${\bf w}_i^T {\bf v}$ is close to zero,  one of the two exponential terms in Eq.~(\ref{eq:Ew}) always dominates such that ${\mathcal E}_{\boldsymbol W}({\bf v}) \approx  \widetilde{{\mathcal E}}_{\boldsymbol W}({\bf v}) $ where
\begin{equation}
\label{eq:Ewap}
\begin{aligned}
 \widetilde{{\mathcal E}}_{\boldsymbol W}({\bf v}) 
= - \sum\limits_{i=1}^{n_h}  \left|{\bf w}_i^T {\bf v} \right| =  - \sum\limits_{i=1}^{n_h}  \left| \sum\limits_{j=1}^{n_v}  W_{ij} v_j \right|.
\end{aligned}
\end{equation}
Eq.~(\ref{eq:Ewap}) can further be approximated by setting ${\bf v} = {\bf 1}$ with all $v_j = +1$,  i.e. $\widetilde{{\mathcal E}}_{\boldsymbol W}({\bf v}) \approx  \widetilde{{\mathcal E}}_{\boldsymbol W}({\bf 1}) $ with
\begin{equation}
\begin{aligned}
 \widetilde{{\mathcal E}}_{\boldsymbol W}({\bf 1}) 
= - \sum\limits_{i=1}^{n_h}  \left| {\rm sum}({\bf w}_i^T ) \right|
= - \sum\limits_{i=1}^{n_h}  \left| \sum\limits_{j=1}^{n_v}  W_{ij} \right|.
\end{aligned}
\end{equation}
In summary, $ {\mathcal E}_{\boldsymbol W}({\bf v})$, $ \widetilde{{\mathcal E}}_{\boldsymbol W}({\bf v})$ and $\widetilde{{\mathcal E}}_{\boldsymbol W}({\bf 1})$ are all good approximations to the original ${\mathcal E}_{\boldsymbol \theta}({\bf v})  $  (Fig.~\ref{fig:combined_rbm_E}a). The increase of mean $ \langle {\mathcal E}_{\boldsymbol \theta}({\bf v}) \rangle $ with temperature coincides with the increase of   $-\left| {\rm sum}({\bf w}_i^T ) \right|$ with temperature, which is evident from Fig.~\ref{fig:ising2d_W_sample}b. At fixed temperature, the decrease of $ \langle {\mathcal E}_{\boldsymbol \theta}({\bf v}) \rangle $  with $n_h$ is  a consequence of the sum $\sum\limits_{i=1}^{n_h} $ in the definition of visible energy.
The variance  $\langle  {\mathcal E}_{\boldsymbol \theta}^2 \rangle - \langle  {\mathcal E}_{\boldsymbol \theta}\rangle^2$ is a useful quantity for phase transition detection, because it reflects the fluctuation of the probability $p_{\boldsymbol \theta}({\bf v})$. In both low $T$ ferromagnetic and high $T$ paramagnetic regimes, $p_{\boldsymbol \theta}({\bf v})$ is relatively homogeneous among different states. When $T$ is close to $T_c$, the variance of $p_{\boldsymbol \theta}({\bf v})$ and ${\mathcal E}_{\boldsymbol \theta}({\bf v})$ is expected to peak (Fig.~\ref{fig:combined_rbm_E}d-e). The abnormal rounded (and even shifted) peaks at large $n_h$ could be a sign of overfitting.

For the all temperature $\cup T$-RBM, the Ising phase transition can be revealed by either the sharp increase of the mean  $ \langle {\mathcal E}_{\boldsymbol \theta}({\bf v}) \rangle $  or the peak of the variance $\langle  {\mathcal E}_{\boldsymbol \theta}^2 \rangle - \langle  {\mathcal E}_{\boldsymbol \theta}\rangle^2$  (Fig.~\ref{fig:combined_rbm_E}c,f). However, this apparent detection can be a trivial consequence of the special composition of the dataset $D_{\cup T}$, which contains Ising configurations at different temperatures in equal proportion. Only configurations at a specific $T$ are fed into the model to calculate  the average quantity at that $T$. Technically, a visible state ${\bf v}$ in $D_{\cup T}$ is not subject to the Boltzmann distribution at any specific temperature. Instead, the true ensemble of $D_{\cup T}$ is a collection of $n_T$ different Boltzmann distributed subsets. Many replicas of the same or similar ferromagnetic states are in $D_{\cup T}$, giving rise to a large multiplicity, high probability and low visible energy for such states. In comparison, high temperature paramagnetic states are all different from each other, and therefore have low $p_{\boldsymbol \theta}({\bf v})$ (high $ {\mathcal E}_{\boldsymbol \theta}({\bf v})$) for each one of them. Knowing this caveat, one should be cautious when monitoring the visible energy of a $\cup T$-RBM to detect phase transition, because changing the proportion of Ising configurations at different temperatures in $D_{\cup T}$ can modify the relative probability of each state.

\subsection{Pseudo-likelihood and entropy estimation}
The likelihood $ {\mathcal L}({\boldsymbol \theta})$ defined in Eq.~(\ref{eq:L}) is conceptually equivalent to the physical entropy $S$  defined by the Gibbs entropy formula, apart from the Boltzmann constant $k_B$ difference (Appendix~\ref{ap:ising}). However, just as entropy $S$ cannot be directly sampled, the exact value of $ {\mathcal L}({\boldsymbol \theta})$ is not accessible. In order to estimate $S$, we calculate the pseudo-likelihood  $\widetilde{\mathcal L}({\boldsymbol \theta})$ instead, which is based on the mean-field like approximation $ p_{\boldsymbol \theta}({\bf v}) \approx \prod_{i=1}^{n_v} p_{\boldsymbol \theta} (v_i | v_{j\ne i})$.  
A similar idea to estimate  free energy was put forward using variational autoregressive networks~\cite{wu2019}.
The true and estimated entropy of $2d$ and $3d$ Ising models using  $T$-RBMs with different  $n_h$ are shown in Fig.~\ref{fig:ising_L} (a-b).
As a comparison, we also consider a ``pseudo-entropy'' with the similar approximation 
\begin{equation}
\widetilde{S} = - k_B \left\langle \sum_{i=1}^{N} p_{T} (s_i | s_{j\ne i}) \right\rangle_{{\bf s}\sim p_T} \approx S
\end{equation}
where the conditional probability
\begin{equation}
p_{T} (s_i | s_{j\ne i}) =  \frac{ e^{- \frac{{\mathcal H}({\bf s})}{k_B T}}   }{   e^{- \frac{{\mathcal H}({\bf s})}{k_B T}}   +   e^{- \frac{{\mathcal H}( [s_1,  \cdots, -s_i, \cdots, s_{N}])}{k_B T}}}
\end{equation}
and the ensemble average  $ \left\langle \cdots \right\rangle_{{\bf s}\sim p_T} $ is taken over states obtained from  Monte Carlo sampling. In both $2d$ and $3d$, $\widetilde{S}$ is lower than the true $S$, especially at high $T$, because a mean-field treatment tends to underestimate fluctuations.

While increasing model complexity by adding hidden units is usually believed  to reduce the reconstruction error, e.g. of energy and heat capacity~\cite{torlai2016,morningstar2018} (see also Supplemental Material), recent study suggests that a trade-off could exist between the accuracy of different statistical quantities~\cite{yevick2021}. Here we find that the pseudo-likelihood of $T$-RBMs with the fewest hidden units in our trials ($n_h=400$) appears to provide the best prediction for entropy. Increasing $n_h$ leads to larger deviations from the true $S$ at higher $T$.  The decreasing of  $\widetilde{\mathcal L}$ with $n_h$ at fixed temperature agrees with the trend of the visible energy. A lower ${\mathcal E}_{\boldsymbol \theta}({\bf v})$ corresponds to a higher $p_{\boldsymbol \theta}({\bf v}) $ and thus a lower $\widetilde{\mathcal L}$ according to its definition.
The surprisingly good performance of $\widetilde{\mathcal L}$  in approximating $S$ could be due to the fact that visible units $v_i$ in RBMs are only indirectly correlated through hidden units, which collectively serve as an effective mean-field on each visible unit. 
We also calculate  $\widetilde{\mathcal L}({\boldsymbol \theta})$ with the all temperature  $\cup T$-RBM in $2d$ (Fig.~\ref{fig:ising_L}c).  Compared with single temperature $T$-RBMs of the same $n_h$ (Fig.~\ref{fig:ising_L}a), the $\cup T$-RBM predicts higher $\widetilde{\mathcal L}({\boldsymbol \theta})$  with considerable deviations even at low $T$. The trend of $\widetilde{\mathcal L}({\boldsymbol \theta})$ also agrees with that of $\langle {\mathcal E}_{\boldsymbol \theta}({\bf v}) \rangle $ (Fig.~\ref{fig:combined_rbm_E}c).

A knowledge about the entropy allows us to estimate the phase transition point according to the thermodynamic relation $C_V = T \frac{d S}{dT}$. We construct this estimated $C_V$ as a function of temperature using $\widetilde{\mathcal L}({\boldsymbol \theta})$ and its numerical fitting, whose peaks are expected to be located at $T_c$ (Supplemental Material). The predicted $T_c$ are compared with the results from entropy and pseudo-entropy, as well as the known exact values in Table~\ref{table:Tc}. It can be seen that single temperature $T$-RBMs capture the transition point fairly well within an error about $1$-$3\%$.
\begin{table}[H]
\setlength\extrarowheight{4pt}
\caption{$T_c$ estimated according to the peak of $T \frac{d \widetilde{\mathcal L} }{dT}$  obtained from single temperature $T$-RBMs and the all temperature $\cup T$-RBM with different number ($n_h$) of hidden units. Predictions from  numerical derivatives $T \frac{d S }{dT} $ and  $T \frac{d \widetilde{S} }{dT} $ are also shown for  comparison.}
\begin{tabular}{cccccccc}
\hline 
model & $ n_h=400$&  $  900$&  $ 1600$&  $  2500$&  $ S$&  $ \widetilde{S}$ & exact \\
\hline 
$2d$ $T$-RBM  & 2.240 & 2.291& 2.316& 2.367& 2.267& 2.367& 2.269\\
$2d$ $\cup T$-RBM & 2.189 & 2.163 & 2.214& - & 2.267& 2.367& 2.269 \\
$3d$ $T$-RBM & 4.444 & 4.434 & 4.444 & - & 4.390& 4.383& 4.511\\
\hline
\end{tabular}
\label{table:Tc}
\end{table} 

\section{Conclusion}
In this work, we trained RBMs using  equilibrium Ising configurations in $2d$ and $3d$ collected from Monte Carlo simulations at various temperatures.  For single temperature $T$-RBMs,  the filters (row vectors) and the inverse filters (column vectors) of the weight matrix exhibit different characteristic patterns and  correlations, respectively, below, around and above the phase transition. These metrics, such as filter sum fluctuation and inverse filter correlation, can be used to locate the phase transition point. The hidden layer ${\bf h}$ on average contains an equal number of $+1$ and $-1$ units, whose variance  decreases as more hidden units are added. The sign of a particular hidden unit $h_i$ is determined by the signs of the filter sum ${\rm sum}({\bf w}_i^T )$ and the magnetization $m$ of the visible pattern. But there is no spatial pattern in the sequence of  positive and negative units in a hidden encoding.  

The visible energy reflects the relative probability  of visible states in the  Boltzmann distribution. Although the mean of visible energy is not directly related to the (physical) internal energy and does not reveal a clear transition, its fluctuation which peaks at the critical point can be used to identify the phase transition. The value and trend of the visible energy can be understood from its several approximation forms, in particular, the sum of the absolute value of filter sums.  The pseudo-likelihood of RBMs is conceptually related to and can be used to estimate the physical entropy.  Numerical differentiation of pseudo-likelihood provides another estimator of the transition temperature because it provides an estimate of the heat capacity. All these predictions about the critical temperature are made by unsupervised RBM learning, for which human labeling of phase types are not needed. 
 
 As a comparison, we also trained an all temperature $\cup T$-RBM whose dataset is a mixture of Boltzmann-distributed states over a range of temperatures. Each filter of this $\cup T$-RBM is featured by one large domain in its receptive field. Although the visible energy and pseudo-likelihood of the  $\cup T$-RBM show certain signature of the phase transition, one should be cautious that this detection could be an artifact of the composition of the dataset. Changing the proportions of Ising configurations at different temperatures could bias the probability and the transition learned by the $\cup T$-RBM.

 By extracting the underlying (Boltzmann) distribution of input data, RBMs capture the rapid (phase) transition of such distribution as the tuning parameter (temperature) is changed,  without knowledge of the physical Hamiltonian. Information about the distribution is completely embedded in the configurations and their frequencies in the dateset.  It would be interesting to see if such a general scheme of RBM learning can be extended to study other physical models of phase transition.

\begin{acknowledgments}
We thank the Duke Kunshan startup funding and the Summer Research Scholars (SRS) program for supporting this work.
\end{acknowledgments}

\appendix

\section{Statistical thermodynamics of Ising model}
\label{ap:ising}
In this appendix, we review the statistical thermodynamics of the Ising model covered in this work.
The internal energy at a given temperature
\begin{align}
\langle E \rangle =   \sum\limits_{ {\bf s}} p_T({\bf s}) {\mathcal H}({\bf s})  = \frac{ \sum\limits_{ {\bf s}}  {\mathcal H}({\bf s})  e^{- \frac{{\mathcal H}\left({\bf s}\right)  }{ k_B T}}     }{ Z_T}  
\end{align}
where $\langle \cdots \rangle$ means to take thermal average over equilibrated configurations. The heat capacity is
 \begin{align}
C_V =  k_B \beta^2  \left(  \langle E^2\rangle   - \langle E \rangle^2  \right)
\end{align}
where $\beta = \frac{1}{k_B T}$ and the heat capacity per spin (or specific heat) is $c_V = C_V/N$.
The magnetization per spin
 \begin{align}
\langle m \rangle = \frac{1}{N} \left \langle  \sum_{i=1}^N s_i  \right \rangle.
\end{align}
In small finite systems, because flips from $m$ to $-m$ configurations are common, we need to take   absolute value $|m|$ before thermal average
  \begin{align}
\langle |m| \rangle = \frac{1}{N} \left \langle   \left| \sum_{i=1}^N s_i  \right| \right \rangle.
\end{align}
The physical entropy can be defined using the Gibbs entropy formula
  \begin{align}
S = -k_B \langle \ln p_T({\bf s}) \rangle = -k_B \sum\limits_{ {\bf s}}  p_T({\bf s})  \ln p_T({\bf s}).
\end{align}

For $2d$ Ising model, the critical temperature  solved from   $\sinh \left(2 \frac{J}{k_B T_c} \right) = 1$ is
$k_B T_c  = \frac{2 J}{\ln (1 + \sqrt{2})} = 2.269185 J.$
Define
\begin{align*}
 K = &\frac{J}{k_{B}T}, ~~~ x=e^{-2K},~~~
 q(K)=\frac{2\sinh2K}{\cosh^{2}2K}\\
 &K_{1}(q)=\int_{0}^{\pi/2}\frac{d\phi}{\sqrt{1-q^2\sin^{2}\phi}}\\
 &E_{1}(q)=\int_{0}^{\pi/2}d\phi\sqrt{1-q^2\sin^{2}\phi},
\end{align*}
analytical results about $2d$ Ising model are expressed as: magnetization per spin~\cite{yang:1952}
\begin{align}
\langle m \rangle & =\left[\frac{1+x^{2}}{(1-x^2)^2}\left(1-6x^{2}+x^{4}\right)^{\frac{1}{2}}\right]^{\frac{1}{4}} \\
&= [1 - \sinh^{-4}(2K)]^{1/8},
\end{align}
internal energy per spin~\cite{kramers:1941}
\begin{equation}
\frac{ \langle E\rangle }{N}=-J\coth 2K \left[1+\frac{2}{\pi}\left(2 \tanh^2
2K-1\right)K_{1}(q)\right],
\end{equation}
specific heat~\cite{plischoke:1994}
\begin{equation}
\begin{aligned}
&c_V = k_B \frac{4}{\pi}\left(K\coth2K\right)^2 \left\{K_{1}(q)-E_{1}(q) - \right.\\
&\left. \left(1-\tanh^{2}2K\right)\left[\frac{\pi}{2}+\left(2\tanh^{2}2K-1\right)K_{1}(q)\right] \right\},
\end{aligned}
\end{equation}
 and the partition function per spin (or free energy per spin $f=F/N$)~\cite{onsager1944} 
\begin{equation}
\begin{aligned}
-\beta f 
 &= \ln (\sqrt{2} \cosh 2K) \\
 &+ \frac{1}{\pi} \int_0^{\pi/2} \ln   \left(1 + \sqrt{1 - q^2 \sin^2 \phi} \right)  d\phi. 
\end{aligned}
\end{equation}
 The equation for entropy can be obtained from thermodynamic relation $F = \langle E \rangle - TS$.

 For $3d$ Ising model, $\langle m \rangle$, $\langle E\rangle $ and $c_V$ can be calculated directly from Monte Carlo sampling~\cite{landau2021}.  The numerical prediction for the critical temperature is $T_c \approx 4.511 \frac{J}{k_B}$~\cite{ferrenberg1991}. Special techniques are needed to compute free energy or entropy. We use the thermodynamic integration in the high temperature regime
 \begin{align}
\begin{aligned}
F & = - N k_B T \ln 2 +  k_B T  \int_{ 0 }^{\frac{1}{k_BT} } \langle  E \rangle d\beta' 
\end{aligned}
\end{align}
or 
\begin{align}
\begin{aligned}
S(T) = \int_0^T \frac{C_V(T')}{T'} dT'
\end{aligned}
\end{align}
in the low temperature regime, since $S(T\to 0)  =0$ and $C_V(T\to 0) \to 0$ for the Ising model.

\section{Energy and probability of RBMs}
\label{ap:rbm}

In this appendix, we review the derivations about the energy and probability of RBMs, which can be found in standard machine learning literature~\cite{murphy2012}.
The visible energy $ {\mathcal E}_{\boldsymbol \theta}({\bf v})$
\begin{widetext}
\begin{equation*}
\begin{aligned}
 {\mathcal E}_{\boldsymbol \theta}({\bf v}) &= - \ln \sum_{\bf h} e^{- E_{\boldsymbol \theta}({\bf v}, {\bf h})} = - \ln p_{\boldsymbol \theta}({\bf v}) - \ln Z_{\boldsymbol \theta}
 =  - \ln  \left[ e^{ \sum\limits_j^{n_v} b_j v_j }  \sum\limits_{\bf h} e^{ \sum\limits_i^{n_h}  \left(\sum\limits_j^{n_v} W_{ij}  v_j + c_i  \right) h_i   }  \right] \\
  &=   - \sum\limits_j^{n_v} b_j  v_j - \ln \left[  \sum\limits_{h_1=-1}^{+1} \sum\limits_{h_2=-1}^{+1} \cdots \sum\limits_{h_{n_h}=-1}^{+1}  \prod\limits_{i=1}^{n_h}e^{   \left(\sum\limits_j^{n_v} W_{ij}  v_j + c_i  \right) h_i   } \right]  
= - \sum\limits_j^{n_v} b_j  v_j  - \ln \left[ \prod\limits_{i=1}^{n_h} \sum\limits_{h_i=-1,1}e^{   \left(\sum\limits_j^{n_v} W_{ij}  v_j + c_i  \right) h_i   } \right] \\
&= - \sum\limits_j^{n_v} b_j  v_j  - \ln \prod\limits_{i=1}^{n_h} \left( e^{   - \sum\limits_j^{n_v} W_{ij}  v_j - c_i }   +  e^{    \sum\limits_j^{n_v} W_{ij}  v_j + c_i     } \right) 
= - \sum\limits_j^{n_v} b_j  v_j  - \sum\limits_{i=1}^{n_h}  \ln \left( e^{   - \sum\limits_j^{n_v} W_{ij}  v_j - c_i } +  e^{    \sum\limits_j^{n_v} W_{ij}  v_j + c_i     } \right)\\
&= -{\bf b}^T {\bf v} - \sum\limits_{i=1}^{n_h}  \ln \left( e^{   - {\bf w}_i^T {\bf v} - c_i  }  +  e^{    {\bf w}_i^T {\bf v} + c_i     } \right).
\end{aligned}
\end{equation*}
\end{widetext}
 
 The conditional probability  
 \begin{equation*}
\begin{aligned}
p_{\boldsymbol \theta}({\bf h}|{\bf v}) & = \frac{   p_{\boldsymbol \theta}({\bf v},{\bf h}) }{p_{\boldsymbol \theta}({\bf v}) } = \frac{ e^{-E_{\boldsymbol \theta}({\bf v}, {\bf h})}}{ e^{-{\mathcal E}_{\boldsymbol \theta}({\bf v})}} 
= \frac{e^{{\bf b}^T {\bf v} }}{e^{-{\mathcal E}_{\boldsymbol \theta}({\bf v})}} e^{ {\bf c}^T {\bf h} + {\bf h}^T{\bf W} {\bf v}} 
\end{aligned}
\end{equation*}
 \begin{equation*}
 =\frac{1}{\Omega_{\boldsymbol \theta}({\bf v})} e^{ {\bf c}^T {\bf h} + {\bf h}^T{\bf W} {\bf v}} 
 \end{equation*}
where the ${\bf h}$-independent constant $\Omega_{\boldsymbol \theta}({\bf v}) = e^{ - {\bf b}^T {\bf v} -{\mathcal E}_{\boldsymbol \theta}({\bf v})} = \sum\limits_{\bf h} e^{ {\bf c}^T {\bf h} + {\bf h}^T{\bf W} {\bf v}}  $ such that $Z_{\boldsymbol \theta} =    \sum\limits_{\bf v}  \Omega_{\boldsymbol \theta}({\bf v})    e^{  {\bf b}^T {\bf v}}  $. So
 \begin{equation*}
\begin{aligned}
p_{\boldsymbol \theta}({\bf h}|{\bf v}) & =  \frac{1}{\Omega_{\boldsymbol \theta}({\bf v})} e^{ \sum\limits_{i=1}^{n_h} c_i h_i + \sum\limits_{i=1}^{n_h}  h_i {\bf w}_{i}^T {\bf v}}  \\
&=  \frac{1}{\Omega_{\boldsymbol \theta}({\bf v})} e^{ \sum\limits_{i=1}^{n_h}  h_i \left( c_i +  {\bf w}_{i}^T {\bf v} \right) } \\
&= \frac{1}{\Omega_{\boldsymbol \theta}({\bf v})} \prod\limits_{i=1}^{n_h} e^{   h_i \left( c_i +  {\bf w}_{i}^T {\bf v} \right) } = \prod\limits_{i=1}^{n_h} p_{\boldsymbol \theta}(h_i | {\bf v})
\end{aligned}
\end{equation*}
from which it can be recognized that $p_{\boldsymbol \theta}(h_i | {\bf v})  \propto e^{   h_i \left( c_i +  {\bf w}_{i}^T {\bf v} \right) }$. The single unit conditional probability 
 \begin{equation}
\begin{aligned}
p_{\boldsymbol \theta}(h_i = 1 | {\bf v}) &= \frac{p_{\boldsymbol \theta}(h_i = 1 | {\bf v})}{p_{\boldsymbol \theta}(h_i = -1 | {\bf v}) + p_{\boldsymbol \theta}(h_i = 1 | {\bf v})} \\
&= \frac{ e^{     c_i +  {\bf w}_{i}^T {\bf v}   } }{e^{    -c_i - {\bf w}_{i}^T {\bf v}  } + e^{    c_i +  {\bf w}_{i}^T {\bf v}   }} \\
& = \frac{1}{1 + e^{ -2(    c_i +  {\bf w}_{i}^T {\bf v}  ) } }\\
&= \sigma\left( 2(c_i +  {\bf w}_{i}^T {\bf v}  ) \right).
\end{aligned}
\end{equation}
Other relations about $p_{\boldsymbol \theta}(h_i = -1 | {\bf v}) $, $p_{\boldsymbol \theta}(v_j = 1 | {\bf h})$ and $p_{\boldsymbol \theta}(v_j = -1 | {\bf h})$ can be found similarly.

\section{Maximum likelihood estimation and gradient descent of RBMs}
\label{ap:rbm_train}
In this appendix, we review the gradient descent algorithm of RBMs derived from  maximum likelihood estimation~\cite{murphy2012}.
The likelihood function for a given dataset $D = [ {\bf v}_1, {\bf v}_2, \cdots, {\bf v}_M ]^T$ is $P_{\boldsymbol \theta}(D)  =  \prod\limits_{m=1}^M  p_{\boldsymbol \theta}({\bf v}_m)$ and maximum likelihood is equivalent to minimum negative log likelihood (or its average)
\begin{equation}
\begin{aligned}
{\boldsymbol \theta}^* &= \arg \max\limits_{\boldsymbol \theta} \prod\limits_{m=1}^M  p_{\boldsymbol \theta}({\bf v}_m) \\
&=\arg \min\limits_{\boldsymbol \theta} \left\{ -\sum\limits_{m=1}^M  \ln p_{\boldsymbol \theta}({\bf v}_m) \right\}  \\
&= \arg \min\limits_{\boldsymbol \theta} \left\{ - \frac{1}{M}\sum\limits_{m=1}^M  \ln p_{\boldsymbol \theta}({\bf v}_m) \right\} \\
&= \arg \min\limits_{\boldsymbol \theta}\langle -\ln p_{\boldsymbol \theta}({\bf v}) \rangle_{{\bf v}\sim p_D} =  \arg \min\limits_{\boldsymbol \theta } {\mathcal L}({\boldsymbol \theta})
\end{aligned}
\end{equation}
where  ${\bf v}\sim p_D$ means to randomly draw ${\bf v}$ from $p_D$ and $\langle \cdots\rangle$ is the expectation value (subject to the distribution). Alternatively, this can be considered as to minimize the Kullbach-Leibler (KL) divergence
\begin{equation*}
\begin{aligned}
 D_{\rm KL}(p_D | p_\theta) &= \sum\limits_{m=1}^M p_D({\bf v}_m) \ln \frac{p_D({\bf v}_m)}{p_\theta({\bf v}_m)} \\
&=\left \langle  \ln  p_D({\bf v}) -\ln  p_\theta({\bf v})\right \rangle_{{\bf v}\sim p_D} \ge 0
\end{aligned}
\end{equation*}
with respect to ${\boldsymbol \theta}$, where only the second term $\left \langle  -\ln  p_\theta({\bf v})\right \rangle_{{\bf v}\sim p_D}  $ depends on parameter ${\boldsymbol \theta}$. In this work, we use  ${\mathcal L}({\boldsymbol \theta})$ as the loss function to train RBMs.

It is sometimes useful to directly monitor the reconstruction error by comparing the input (${\bf v}$) and reconstructed configurations (${\bf v}'$), or more quantitatively, by the (normalized) cross entropy 
\begin{equation}
\begin{aligned}
 {\rm CE} &= \left\langle -\frac{1}{n_v} \sum_{j=1}^{n_v}   \left[ \mathbbm{1}_{v_j = +1} \ln  p_{\boldsymbol \theta}(  v_j' = +1  | {\bf h}) \right.\right.\\
 & +  \left.\left.\mathbbm{1}_{v_j = -1} \ln  p_{\boldsymbol \theta}(  v_j' = -1  | {\bf h}) \right] \right \rangle_{{\bf v}\sim p_D}
\end{aligned}
\end{equation}
where  the indicator function $ \mathbbm{1}_A = 1$ if $A$ is true, or $0$ if  $A$ is false. 

The gradient of the loss function
\begin{equation}
\begin{aligned}
 \nabla_{\boldsymbol \theta}  {\mathcal L}({\boldsymbol \theta})  & =  \nabla_{\boldsymbol \theta} \langle  {\mathcal E}_{\boldsymbol \theta}({\bf v})  \rangle_{{\bf v}\sim p_D}  + \nabla_{\boldsymbol \theta} \ln Z_{\boldsymbol \theta} \\
 &=   \langle  \nabla_{\boldsymbol \theta} {\mathcal E}_{\boldsymbol \theta}({\bf v})  \rangle_{{\bf v}\sim p_D}  + \nabla_{\boldsymbol \theta} \ln Z_{\boldsymbol \theta} 
 \end{aligned}
\end{equation}
where
\begin{equation*}
\begin{aligned}
\nabla_{\boldsymbol \theta} \ln Z_{\boldsymbol \theta} &= \frac{ \nabla_{\boldsymbol \theta} Z_{\boldsymbol \theta} }{Z_{\boldsymbol \theta}} =  \frac{ \nabla_{\boldsymbol \theta}  \sum\limits_{{\bf v} } e^{- {\mathcal E}_{\boldsymbol \theta}({\bf v})} }{Z_{\boldsymbol \theta}}  \\
&=  \frac{  \sum\limits_{{\bf v} } \nabla_{\boldsymbol \theta}  e^{- {\mathcal E}_{\boldsymbol \theta}({\bf v})} }{Z_{\boldsymbol \theta}} 
= - \frac{  \sum\limits_{{\bf v} }  e^{- {\mathcal E}_{\boldsymbol \theta}({\bf v})} \nabla_{\boldsymbol \theta}  {\mathcal E}_{\boldsymbol \theta}({\bf v}) }{Z_{\boldsymbol \theta}}  \\
&=  - \sum\limits_{{\bf v} }   p_{\boldsymbol \theta}({\bf v}) \nabla_{\boldsymbol \theta}  {\mathcal E}_{\boldsymbol \theta}({\bf v}) \\
&= - \langle  \nabla_{\boldsymbol \theta}  {\mathcal E}_{\boldsymbol \theta}({\bf v})  \rangle_{{\bf v}\sim p_\theta}.
 \end{aligned}
\end{equation*}
So,
\begin{equation}
\begin{aligned}
 \nabla_{\boldsymbol \theta}   {\mathcal L}({\boldsymbol \theta}) &= \langle  \nabla_{\boldsymbol \theta}  {\mathcal E}_{\boldsymbol \theta}({\bf v})  \rangle_{{\bf v}\sim p_D}  - \langle  \nabla_{\boldsymbol \theta}  {\mathcal E}_{\boldsymbol \theta}({\bf v})  \rangle_{{\bf v}\sim p_\theta} \\
 &= {\rm positive~phase} + {\rm negative~phase} \\
 & = {\rm data~term } + {\rm model~term}
  \end{aligned}
\end{equation}
In both positive and negative phase,
\begin{equation*}
 \nabla_{\boldsymbol \theta} {\mathcal E}_{\boldsymbol \theta}({\bf v}) = 
 \nabla_{\boldsymbol \theta} \left[
  -{\bf b}^T {\bf v} - \sum\limits_{i=1}^{n_h}  \ln \left( e^{   - {\bf w}_i^T {\bf v}  -c_i  } +  e^{    {\bf w}_i^T {\bf v} + c_i     } \right) \right]
\end{equation*}
which has components
\begin{equation}
\begin{aligned}
\frac{\partial {\mathcal E}_{\boldsymbol \theta}({\bf v}) }{\partial W_{ij}} &= -\frac{- v_j e^{    -{\bf w}_i^T {\bf v} - c_i }  + v_j e^{    {\bf w}_i^T {\bf v} + c_i } }{ e^{   - {\bf w}_i^T {\bf v}  -c_i  } +  e^{    {\bf w}_i^T {\bf v} + c_i }}  \\
& = v_j  \frac{e^{    -{\bf w}_i^T {\bf v} - c_i }  - e^{    {\bf w}_i^T {\bf v} + c_i } }{ e^{   - {\bf w}_i^T {\bf v}  -c_i  } +  e^{    {\bf w}_i^T {\bf v} + c_i }}  = - v_j \tanh \left(    {\bf w}_i^T {\bf v} + c_i \right) \\
&= -v_j \left[ (-1)p_{\boldsymbol \theta}(h_i=-1 | {\bf v}) + (+1) p_{\boldsymbol \theta}(h_i=1 | {\bf v})  \right]  \\
&=  - v_j  \langle h_i \rangle_{h_i \sim  p_{\boldsymbol \theta}(h_i| {\bf v})} \\
\frac{\partial {\mathcal E}_{\boldsymbol \theta}({\bf v}) }{\partial c_i} &= -\frac{-e^{    -{\bf w}_i^T {\bf v} - c_i } +  e^{    {\bf w}_i^T {\bf v} + c_i } }{ e^{    -{\bf w}_i^T {\bf v} - c_i }  +  e^{    {\bf w}_i^T {\bf v} + c_i }} =  -  \tanh \left(    {\bf w}_i^T {\bf v} + c_i \right)  \\
&=  - \left[ (-1)p_{\boldsymbol \theta}(h_i=-1 | {\bf v}) + (+1) p_{\boldsymbol \theta}(h_i=1 | {\bf v})  \right] \\
&= - \langle h_i \rangle_{h_i \sim  p_{\boldsymbol \theta}(h_i| {\bf v})}\\
 \frac{\partial {\mathcal E}_{\boldsymbol \theta}({\bf v}) }{\partial b_j} &= - v_j.
 \end{aligned}
\end{equation}

To evaluate the expectation value $ \langle \nabla_{\boldsymbol \theta} {\mathcal E}_{\boldsymbol \theta}({\bf v}) \rangle$,  in positive phase   ${\bf v}$  can be directly drawn from the dataset,   while in negative phase ${\bf v}$ must be sampled from  the model distribution $p_{\boldsymbol \theta}({\bf v})$.
In practice, as an approximation, Markov chain Monte Carlo (MCMC) method is used to generate ${\bf v}$ states that obey the distribution $p_{\boldsymbol \theta}({\bf v})$, such that
\begin{equation}
\langle  \nabla_{\boldsymbol \theta}  {\mathcal E}_{\boldsymbol \theta}({\bf v})  \rangle_{{\bf v}\sim p_\theta} \approx \frac{1}{\rm sample ~ size}\sum\limits_{{\bf v} \sim p_\theta}  \nabla_{\boldsymbol \theta}  {\mathcal E}_{\boldsymbol \theta}({\bf v}).
\end{equation}

Using the conditional probability, $p_{\boldsymbol \theta}({\bf h} | {\bf v})$ and $p_{\boldsymbol \theta}({\bf v} | {\bf h})$, we can generate a sequence of states
$${\bf v}^{(0)} \to {\bf h}^{(0)} \to {\bf v}^{(1)} \to {\bf h}^{(1)} \to \cdots  \to {\bf v}^{(t)} \to {\bf h}^{(t)} \to \cdots.$$
As $t \to \infty$, the MCMC converges with $({\bf v}^{(t)}, {\bf h}^{(t)}) \sim p_{\boldsymbol \theta}({\bf v}, {\bf h})$  and ${\bf v}^{(t)} \sim p_{\boldsymbol \theta}({\bf v})$.

Markov chain starting from a random ${\bf v}^{(0)} $ takes a lot of steps to equilibrate. There are two ways to speed up the sampling~\cite{hinton2012}
\begin{itemize}
\item $k$ step contrastive divergence (CD-$k$)

For each parameter update, draw ${\bf v}^{(0)}$ (or a minibatch) from the training data $D = [ {\bf v}_1, {\bf v}_2, \cdots, {\bf v}_M ]^T$ and run Gibbs sampling for $k$ steps. Even CD-1 can work reasonably well.

\item persistent contrastive divergence (PCD-$k$)

Always keep the same MC during the entire training process. For each parameter update, run this persistent MC for another $k$ steps to collect ${\bf v}$ states.
\end{itemize}

%

\end{document}